\documentclass[twocolumn,preprintnumbers,amsmath,amssymb,prl,superscriptaddress]{revtex4-1}
\usepackage{graphicx} 
\usepackage{hyperref}
\usepackage{amsfonts}
\usepackage{mathrsfs} 
\usepackage[utf8]{inputenc}
\usepackage{hyperref} 
\usepackage[capitalise]{cleveref}
\usepackage{orcidlink} 
\usepackage{soul}
\usepackage{pgfplots}
\usepackage{tikz-cd}
\usepackage{tikz-3dplot}
\usepackage{lipsum, babel}
\usepackage{multirow}
\usepackage{setspace}
\usepgfplotslibrary{colormaps}
\pgfplotsset{
    compat=newest,
    colormap={mycolormap}{color=(lightgray) color=(white) color=(lightgray)}
}

\tdplotsetmaincoords{70}{0}
\tikzset{declare function={torusx(\u,\v,\R,\r)=cos(\u)*(\R + \r*cos(\v)); 
torusy(\u,\v,\R,\r)=(\R + \r*cos(\v))*sin(\u);
torusz(\u,\v,\R,\r)=\r*sin(\v);
vcrit1(\u,\th)=atan(tan(\th)*sin(\u));
vcrit2(\u,\th)=180+atan(tan(\th)*sin(\u));
disc(\th,\R,\r)=((pow(\r,2)-pow(\R,2))*pow(cot(\th),2)+%
pow(\r,2)*(2+pow(tan(\th),2)))/pow(\R,2);
umax(\th,\R,\r)=ifthenelse(disc(\th,\R,\r)>0,asin(sqrt(abs(disc(\th,\R,\r)))),0);
}}

\definecolor{DESYO}{RGB}{241,143,31}
\definecolor{DESYB}{RGB}{0,159,223}
\definecolor{DESYC}{RGB}{153,0,18}

\newcommand{\SL}{\mathrm{SL}}

\newcommand{\rmO}{\mathrm{O}}
\newcommand{\U}{\mathrm{U}}

\newcommand{\fstop}{.}
\newcommand{\fcomma}{,}
\newcommand{\taub}{\bar{\tau}}

\def\ie{\begin{equation}\begin{aligned}}
\def\fe{\end{aligned}\end{equation}}

\renewcommand{\section}[1]{\bigskip \noindent {\it #1~ }}
\renewcommand{\subsection}[1]{{}}

\begin{document}
\preprint{DESY 23-063}

\title{Duality Origami: Emergent Ensemble Symmetries in Holography and Swampland}

\author{Meer Ashwinkumar\,\orcidlink{0000-0002-4832-3277}}
\affiliation{Kavli IPMU, University of Tokyo, Kashiwa, Chiba 277-8583, Japan}

\author{Jacob M.\ Leedom\,\orcidlink{0000-0003-4911-2188}}
\affiliation{Deutsches Elektronen-Synchrotron DESY, Notkestr.\ 85, 22607 Hamburg, Germany}

\author{Masahito Yamazaki\,\orcidlink{0000-0001-7164-8187}}
\affiliation{Kavli IPMU, University of Tokyo, Kashiwa, Chiba 277-8583, Japan}
\affiliation{Trans-Scale Quantum Science Institute, The University of Tokyo, Tokyo 113-0033, Japan}

\begin{abstract}
We discuss the interrelations between several ideas in quantum gravity -- holography, the
Swampland, and the concept of ensemble averaging. To do so, we study ensemble averages of
Narain-type theories associated with general even quadratic forms and their holographic duals.
We establish the emergence of global symmetries and discuss their consistency with
conjectures forbidding such symmetries. We also discuss how the spectral decomposition of
Narain partition functions suggests a natural embedding of ensemble averaging within the low-energy limit of certain string compactifications, which in turn allows a connection with the
Swampland program.
\end{abstract}

\maketitle 
\section{Introduction: Swampland versus Ensembles}


In recent years two approaches to understanding quantum gravity have emerged. On one hand, there is the Swampland program~\cite{Vafa:2005ui,Ooguri:2006in}, which posits the existence of non-trivial consistency conditions for low-energy effective field theories (EFTs) to be embedded in an ultraviolet (UV) theory of quantum gravity. On the other, there is the holographic approach that leverages dualities between $(D+1)-$dimensional anti-de Sitter (AdS) spacetimes and $D$-dimensional conformal field theories (CFTs).

Standard holographic dualities are argued from the top-down 
in particular string compactifications---the quintessential example being the correspondence between Type IIB on $AdS_5\times S^5$ and $4D$ $\mathcal{N}=4$ Super-Yang Mills theory~\cite{Maldacena:1997re}.  However, there are also bottom-up holographic constructions that utilize ensemble averages of theories. Known examples of this form include Jackiw-Teitelboim (JT) gravity~\cite{Saad:2019lba,Stanford:2019vob} and Narain Ensembles~\cite{Afkhami-Jeddi:2020ezh,Maloney:2020nni}.

Ensemble average dualities are puzzling from both the standard holography and Swampland viewpoints. Indeed, at first glance ensemble averaging appears counter to the one-to-one holographic dualities of the standard approach. Furthermore, ensemble-averaged theories provide loopholes to several Swampland conjectures, including one of the most well-known conjectures excluding 
exact global symmetries in theories of quantum gravity \cite{Misner:1957mt,Banks:1988yz,Kamionkowski:1992mf,Kallosh:1995hi,Banks:2010zn,Harlow:2018jwu,Harlow:2018tng,Harlow:2020bee,Chen:2020ojn,Belin:2020jxr}. Based on these considerations, there have been arguments that ensemble averaging is explicitly excluded by the Swampland program, at least in ten-dimensional string theory \cite{McNamara:2020uza}.  

In this \textit{Letter}, we address the relationship between ensembles, holography, and the Swampland by outlining the physical and mathematical framework to embed Narain ensembles into a more standard approach to holography. We first establish \textit{emergent ensemble symmetries} that are global symmetries in the $3D$ gravity dual of generalized Narain ensembles~\cite{Ashwinkumar:2021kav}~\footnote{See also~\cite{Benini:2022hzx}, and in particular~\cite{Hsin:2020mfa} for discussion on an emergent $O(N)$ symmetry in the SYK model.}. We then relate ensembles to a pre-averaged duality involving Maxwell-Chern-Simons theory and its embedding in string compactifications. This allows us to make contact with several Swampland conjectures. Finally, we utilize the theory of automorphic forms to illustrate that the ensemble average of the generalized Narain CFTs is a coarse-grained description
of Maxwell-Chern-Simons and how additional fluctuations serve to break the emergent ensemble symmetries.

\section{Global Symmetries in Chern-Simons Theories}

We first consider global symmetries in 3D Chern-Simons theories. The relevance of such theories to our discussion is that the bulk duals of the ensemble-averaged CFTs are exotic theories of quantum gravity defined by a particular sum over geometries. This is elucidated in the subsequent section, but for now we note that each geometry supports a theory that appears, \textit{perturbatively}, to be an Abelian $\U(1)^{p+q}$ Chern-Simons theory whose
action is given by \footnote{This theory has been studied intensively in condensed matter physics, see e.g.\ \cite{Lu:2012dt}.}
\begin{equation}
    S_{\rm CS} = \frac{1}{4\pi} \int d^3x \sum_{i,j=1}^{p+q} Q_{ij} A_i dA_j \fstop
\label{eq:bulkCS}
\end{equation}
Here $Q$ is a symmetric, integer-valued matrix of signature $(p,q)$. This is the \textit{level matrix} of the Chern-Simons theory, and a particular level matrix defines a specific ensemble of boundary CFTs, as described below. To define the path integral of~\cref{eq:bulkCS} on an arbitrary 3-dimensional manifold, the diagonal entries of the level matrix must be even integers. For spin manifolds, the diagonal entries need only be integer. 


The theory with classical action \cref{eq:bulkCS} is an Abelian topological quantum field theory (TQFT) whose observables are Wilson lines. These can be thought of as the worldlines of particles with fractional statistics, i.e. anyons.  Abelian TQFTs are completely determined by their anyon data: the anyon fusion algebra and the topological spin of the anyons (and the chiral central charge, which in our case is $p-q$) \cite{Belov:2005ze,Stirling:2008bq,Kapustin:2010hk}, \footnote{In general, a TQFT is determined by the group of anyons and their fusion as well as additional data that defines a modular tensor category. This data includes topological spins, an anyon fusion algebra, a representation of the modular group, and associators and braiding isomorphisms.}. The theory is a \textit{bosonic} (\textit{fermionic}) TQFT if the diagonal entries of $Q$ are even (odd). In this work we consider only bosonic TQFTs, but the ideas can be generalized to fermionic TQFTs utilizing the discussion in~\cite{Ashwinkumar:2021kav}. 

The anyon content for the TQFT defined by~\cref{eq:bulkCS} is determined by $Q$. The number of anyons is simply $\lvert\text{det}(Q)\rvert$, and the anyons can be represented by the integral vectors contained within the parallelepiped formed by the columns of $Q$. Alternatively, the anyons are given by elements of the discriminant group 
$\mathscr{D}$. One can use $Q$ to define an even, integral lattice $\Lambda\cong \mathbb{Z}^{p+q}$. The inner product on $\Lambda$ is defined by $Q$ as $\langle \ell,\ell^\prime\rangle = Q[\ell,\ell^\prime]  =\sum_{i,j=1}^{p+q} Q_{ij}\ell^i \ell'^j$. Note that $Q[\ell,\ell^\prime]\in \mathbb{Z}$ and $Q[\ell] := Q[\ell,\ell] \in  2\mathbb{Z}$ for all $\ell,\ell^\prime\in \Lambda$, as required of an even lattice. The discriminant group of $\Lambda$ is then $\mathscr{D} := \Lambda^* / \Lambda$, where $\Lambda^{*}$ denotes the dual lattice of $\Lambda$~\footnote{The dual lattice of $\Lambda$ is defined as $\Lambda^*:=\{x\in \mathbb{R}^{p+q} |\; Q[x,\ell]\in \mathbb{Z}\;\; \forall \ell\in\Lambda\}$.}. The discriminant group is a finite Abelian group of order $\lvert\mathscr{D}\rvert = \lvert\text{det}(Q)\rvert$ and the anyon fusion algebra is simply defined by addition of two elements of $\mathscr{D}$. The topological spin $\theta$ of an anyon $\alpha\in\mathscr{D}$ 
is given by~\footnote{In an abuse of notation, $\alpha$ is used to denote both an equivalence class in $\mathscr{D}$ as well as an element of $\Lambda^*$ that is a representative of this class.} 
\begin{equation}\label{eq:topospin}
    \theta(\alpha) = \exp\left( i \pi Q[\alpha]\right) \fstop
\end{equation}
 The remaining TQFT data can be derived from $\mathscr{D}$ and $\theta$: the braiding phase of two anyons is 
\begin{equation}\label{eq:braids}
    \mathcal{B}(\alpha,\beta) = \frac{\theta(\alpha+ \beta)}{\theta(\alpha)\theta(\beta)} =\exp\left( 2i \pi Q[\alpha, \beta]\right) \fcomma
\end{equation}
and the S- and T-transformations generating a representation of the modular group are
\begin{align}
\label{eq:ST}
    S_{\alpha\beta} &= \frac{\mathcal{B}(\alpha,\beta)}{\sqrt{\lvert\mathscr{D}\rvert}}\;,
    \quad 
    T_{\alpha\beta} = e^{-\frac{\pi i (p-q)}{12}}\theta(\alpha) \delta_{\alpha\beta} \fstop
\end{align}

Abelian Chern-Simons theories have global one-form and zero-form symmetries
which arise from automorphisms of the TQFT data~\cite{Barkeshli:2014cna,Delmastro:2019vnj} \footnote{For Abelian TQFTs, the 0-form and 1-form symmetries factorize and the 2-group symmetry is the trivial product.}. 
The one-form symmetry group is identical to the discriminant group $\mathscr{D}$ (cf.\ \cite{Hsin:2018vcg}).
The zero-form symmetry group is defined via permutations of the anyons that preserve the anyon data $(\mathscr{D}, \theta)$: \footnote{In this Letter, we consider unitary 
zero-form symmetries. Similar discussions apply when we consider anti-unitary zero-form symmetries, 
which map the topological spins to their complex conjugates.}
\begin{align}\label{zero-form_def}
\mathrm{Aut} (\mathscr{D}, \theta) := \left\{ \sigma \in \textrm{Aut}(\mathscr{D})  \, | \, \theta(\alpha ) = \theta( \sigma\cdot \alpha )  \, \forall \alpha  \in \Lambda \right\} \fstop
\end{align}
Where $\text{Aut}(\mathscr{D})$ denotes automorphisms of the discriminant group (i.e. arbitrary permutations of anyons).

The zero-form global symmetries can be classified as either \textit{classical} or \textit{quantum}. Classical global symmetries arise from the symmetries of the Lagrangian--- for TQFT data $(\mathscr{D}, \theta)$ defined by a quadratic form $Q$, the classical symmetry group is 
\begin{equation}\label{eq:O_Q}
\rmO_Q(p,q;\mathbb{Z}) := \{\Sigma\in \text{GL}(p+q,\mathbb{Z}) \;\; \rvert\;\;  \Sigma^TQ\Sigma = Q \} \fcomma
\end{equation} 
All elements $\Sigma$ of $\rmO_Q(p,q;\mathbb{Z})$ satisfy the criterion in~\cref{zero-form_def}, and re-definitions $A_i\rightarrow \Sigma_{ij}A_j$ make manifest that $\Sigma$ defines a symmetry of the Lagrangian in~\cref{eq:bulkCS}. 

On the other hand, quantum symmetries are not manifest at the level of the Lagrangian but instead satisfy Ward identities.
Note that different quadratic forms $Q, Q'$ can lead to identical anyon data $(\mathscr{D}, \theta)$,
in which case the two Lagrangians associated with $Q, Q'$ are 
quantum-equivalent \cite{Belov:2005ze,Stirling:2008bq,Kapustin:2010hk} \footnote{More precisely, the two theories are equivalent up to invertible TQFTs, whose Hilbert space is one-dimensional.}. Thus it may be that two ensembles associated with distinct quadratic forms lead to identical averages. Along similar lines, whether a symmetry is classical or not is expected to depend on the choice of the classical Lagrangian (i.e., the duality frame), see, e.g.,~\cite{Delmastro:2019vnj}.

\section{Holography of General Narain Ensembles}

To discuss the emergence of global symmetries, 
we must first detail the procedure of ensemble averaging certain classes of 2D CFTs.
We follow the general discussion in Ref.~\cite{Ashwinkumar:2021kav}, which extended the previous arguments of Refs.~\cite{Maloney:2020nni,Afkhami-Jeddi:2020ezh}
(see also Refs.~\cite{Cotler:2020ugk,Cotler:2020hgz,Benjamin:2021wzr,Dong:2021wot,Henriksson:2022dml,AKLY} for variations).

The 2D CFTs we consider are generalizations of standard Narain CFTs. Narain CFTs~\cite{Narain:1985jj,Narain:1986am} are traditionally associated to the worldsheet CFT of toroidal superstring compactifications. The primary operators of Narain CFT consist of U(1) current operators and vertex operators. These vertex operators furnish a lattice -- the \textit{Narain lattice}. This is an even, self-dual lattice defined by a unimodular quadratic form along with continuous parameters that parametrize the Narain moduli space.

A \textit{generalized} Narain CFT differs from a standard Narain CFT in that it is defined with a quadratic form that is not unimodular, and so the generalized Narain lattice is not self-dual. In anticipation of connection to~\cref{eq:bulkCS}, we consider a CFT with its quadratic form given by the $(p+q)\times(p+q)$-dimensional matrix $Q$. The $pq$-dimensional moduli space of the CFT is then the double coset
\begin{equation}
\mathcal{M}_{Q} :=
\mathrm{O}_Q(p,q;\mathbb{Z}) \big\backslash (\rmO(p,q; \mathbb{R})\big/(\rmO(p; \mathbb{R})\times \rmO(q; \mathbb{R})) \fstop
\label{eq:M_Q}
\end{equation}
Note that the T-duality group of the CFT is precisely the classical symmetry group associated to the TQFT defined by $Q$ in~\cref{eq:O_Q}.

Since the generalized Narain lattice is not self-dual, the CFT has a family of partition functions associated with elements $\alpha$ of the discriminant group associated with the lattice $\Lambda$:
\begin{equation}
    Z^{Q}_{\alpha}(\tau, \overline{\tau};m) =
\frac{\vartheta^Q_{\alpha}(\tau;m)}{\eta^p(\tau) \overline{\eta}^q(\overline{\tau})} \fstop
\label{Eq:ZQh}
\end{equation}
Where $\tau$ is the modular parameter of the torus defining the partition function and $m$ collectively denotes the CFT moduli. The theta function $\vartheta^\Lambda_{\alpha}(\tau;m)$ is given by
\begin{equation}
    \begin{aligned}
         \vartheta^Q_{\alpha}(\tau;m) :&= \sum_{\ell \in {\Lambda+\alpha}} e^{i \pi \tau_1 Q[\ell]-  \pi \tau_2 H[\ell]}\\
         &= \sum_{\ell \in {\Lambda+\alpha}} (\theta(\ell))^{\tau_1} e^{-\pi\tau_2H[\ell]}\fstop
    \end{aligned}
  \label{eq:lattheta}
\end{equation}
The dependence on $m$ is encoded in the \textit{Hamiltonian} $H$ -- it is a matrix that depends on $pq$ parameters and satisfies the property $HQ^{-1}H=Q$ (see~\cite{Ashwinkumar:2021kav} for more details). The sum over the lattice $\Lambda$ can be interpreted as a trace over the states related to vertex operators by the usual state-operator correspondence. The last equality in~\cref{eq:lattheta} makes manifest the connection between the generalized CFTs and the Chern-Simons theory of~\cref{eq:bulkCS} as the partition functions are determined by the topological spins $\theta(\alpha)$ of the anyons, up to the contributions from $H$. It is this contribution that gets washed out during the averaging procedure.

Under the generators of $\SL(2,\mathbb{Z})$, the theta functions transform as
\begin{equation}\label{eq:theta_ST}
\begin{split}
  &\vartheta^{Q}_{\alpha }(\tau+1; m) =  
    e^{\frac{\pi i (p-q)}{12}} \sum_{\beta \in  \mathscr{D}} T_{\alpha \beta}  \vartheta^{Q}_{\beta }(\tau; m) \fcomma\\
  &\vartheta^{Q}_{\alpha }\left(-\frac{1}{\tau};  m\right) =
    e^{-\frac{\pi i}{4}(p-q)} \tau^{\frac{p}{2}} \overline{\tau}^{\frac{q}{2}} \sum_{\beta \in  \mathscr{D}}
    S_{\alpha \beta} \vartheta^{Q}_{\beta}(\tau; m) \fcomma
\end{split}
\end{equation}
where $S, T$ are the modular matrices \cref{eq:ST} for the anyons. By repeated use of~\cref{eq:theta_ST}, we can determine the transformation under a general element $M = \begin{pmatrix}a & b\\ c & d\end{pmatrix}\in\SL(2,\mathbb{Z})$,
\begin{equation}\label{eq:theta_Lambda_SL2}
    \vartheta^{Q}_{\alpha}(\tau_{M};m)
      = (c\tau+d)^{\frac{p}{2}} (c\overline{\tau}+d)^{\frac{q}{2}} \sum_{\beta \in \mathscr{D}} \lambda^{Q}_{\alpha, \beta}(M)\,\vartheta^{Q}_{\beta}(\tau;m) 
\end{equation}
where $\tau_M := (a\tau+b)(c\tau+d)^{-1}$ and 
\begin{equation} \label{lambda_def}
 \begin{aligned}
     \lambda^{Q}_{\alpha, \beta}(M) := &\frac{1}{\sqrt{\lvert\mathscr{D}_Q\rvert} } \, e^{-\frac{\pi i}{4}(p-q)} c^{-\frac{p+q}{2}}  \\ &\times \sum_{\ell_c \in \Lambda/( c  \Lambda)} e^{\frac{\pi i}{ c  }\left(a Q[\ell_c +\alpha] -2Q[\ell_c +\alpha, \beta]+ d   Q[\beta] \right)} \fstop
 \end{aligned}
\end{equation}
It is important to note that~\cref{eq:theta_Lambda_SL2} implies that the partition functions in~\cref{Eq:ZQh} are not modular invariant -- indeed, they constitute a vector-valued modular form of $\SL(2,\mathbb{Z})$, and the 2d theory should be considered as a relative theory with respect to the 3d bulk. This should be taken in contrast with typical discussions of 2D CFTs where modular invariance of the partition function is emphasized as part of necessary and sufficient conditions to define the CFT on arbitrary Riemann surfaces~\cite{Moore:1988qv}. 

In case one is interested in strictly modular invariant (i.e.\ absolute) theory, there are several techniques to adapt the theories as defined thus far such that they respect the modular invariance. In cases were the signature of $Q$ satisfies $p-q = 0$ mod $24$, one can construct a modular invariant partition function by summing over a subset of the anyons that close under $\SL(2,\mathbb{Z})$ transformations. From the Chern-Simons point of view, this corresponds to gauging a Lagrangian sub-group of the global 1-form symmetry group of the Abelian CS theory. A similar approach was taken in~\cite{Benini:2022hzx}, but here we are able to perform such a gauging without restricting to rational values of the moduli. Another approach is to consider 
a combinations of two generalized CFTs as
$Z_{inv}  = \sum_{\alpha\in\mathscr{D} } \vartheta^{\widetilde{Q}}_\alpha(\tau;m)\vartheta^Q_\alpha(\tau;m)$ with $\widetilde{Q} = -Q$. This is modular invariant thanks to ~\cref{lambda_def}.

We now consider the ensemble averaging of generalized Narain CFTs. The ensemble average of the partition functions in~\cref{Eq:ZQh} is defined as  
\begin{equation}
\langle \vartheta^Q_{\alpha} \rangle (\tau) := 
\frac{1}{\textrm{Vol}(\mathcal{M}_{Q, \alpha}) } \int_{\mathcal{M}_{Q, \alpha}} [dm] \, \vartheta^Q_{\alpha}(\tau;m) \fcomma
    \label{eq:siegelavg}
\end{equation}
where $[dm]$ denotes the canonical measure associated with the Zamolodchikov metric of the moduli space. The region of integration $\mathcal{M}_{Q, \alpha}$ is a cover of the moduli space $\mathcal{M}_Q= \mathcal{M}_{Q, \alpha=0}$\footnote{Since the label $\alpha$ transforms non-trivially under 
a general element of $\rmO_Q(p,q;\mathbb{Z})$, an isolated theta function $\vartheta_{\alpha}$ in~\cref{eq:lattheta}
is defined over the space $\mathcal{M}_{Q, \alpha} :=
  \mathrm{O}_{Q, \alpha}(p,q;\mathbb{Z}) \big\backslash (\rmO(p,q; \mathbb{R})\big/(\rmO(p; \mathbb{R})\times \rmO(q; \mathbb{R}))$}. For $p+q>4$, this average can be evaluated and yields the Siegel-Eisenstein series~\cite{MR67930,Siegel_Lecture,Maloney:2020nni,Ashwinkumar:2021kav} 
\begin{equation}
   E^{Q}_{\alpha}(\tau, \bar{\tau}) := \delta_{\alpha \in \Lambda} +
 \sum_{(c,d)=1,\, c>0} \, \frac{\gamma^Q_{\alpha}(c,d)}{ (c\tau+d)^{\frac{p}{2}} (c\overline{\tau}+d)^{\frac{q}{2}}}  \fcomma
 \label{eq:E_as_sum}
\end{equation}
where the sum is over integers $c$ and $d$ that are co-prime (i.e. $(c,d)=1$) and $\delta_\alpha$ is equal to unity for $\alpha\in\Lambda$ and zero otherwise. The factor $\gamma^Q_{\alpha}(c,d)$ is defined from the transformation matrix of the theta functions in~\cref{lambda_def} as $\gamma^Q_{\alpha}(c,d):=\lambda^Q_{\alpha, 0}(M^{-1})$.

We can now consider the holographic interpretation of the averaged CFT partition function in~\cref{eq:E_as_sum}. As alluded to above, the bulk dual of the ensemble average is an exotic theory of quantum gravity related to the Abelian CS theory of~\cref{eq:bulkCS} in a manner we now make precise.
The bulk interpretation of~\cref{eq:E_as_sum} involves identifying each term in the sum  as arising from a geometry in the 3D gravitational theory \cite{Afkhami-Jeddi:2020ezh,Maloney:2020nni}:
a pair of co-prime integers defines an ``$\SL(2, \mathbb{Z})$ black hole'' $M_{(c,d)}$ \cite{Maldacena:1998bw,Dijkgraaf:2000fq}, where $M_{(0,1)}$ and  $M_{(1,0)}$ are thermal AdS$_3$ and the BTZ black hole \cite{Banados:1992wn}, respectively. 
 
The size of the contribution from each geometry is determined by partition functions of the Abelian Chern-Simons theory of~\cref{eq:bulkCS}. In particular, the coefficient $\gamma^Q_{\alpha}(c,d)$ coincides with the lens space partition function of the Chern-Simons theory with an insertion of a Wilson line labeled by $\alpha$~\cite{Witten:1988hf,MR1175494}, as pointed out in \cite{Ashwinkumar:2021kav}. This dependence on $\alpha\in\mathscr{D}$ underscores that the bulk dual has objects that appear to be non-trivial anyons whose worldlines wrap the non-contractible cycle of $M_{(c,d)}$. Thus while the true nature of the post-ensemble averaged bulk dual is somewhat mysterious, the CS theory of~\cref{eq:bulkCS} provides a perturbative description on each bulk geometry. The construction of the holographic correspondence is depicted in~\cref{fig:duality}.

\begin{figure}
\centering
\scalebox{0.5}{
\begin{tikzpicture}
\pgfmathsetmacro{\R}{4}
\pgfmathsetmacro{\r}{1}

\begin{scope}[tdplot_main_coords,scale=0.5,xshift=-12cm,yshift=3.8cm]
 \draw[thick,fill=gray,even odd rule,fill opacity=0.2] plot[variable=\x,domain=0:360,smooth,samples=71]
 ({torusx(\x,vcrit1(\x,\tdplotmaintheta),\R,\r)},
 {torusy(\x,vcrit1(\x,\tdplotmaintheta),\R,\r)},
 {torusz(\x,vcrit1(\x,\tdplotmaintheta),\R,\r)}) 
 plot[variable=\x,
 domain={-180+umax(\tdplotmaintheta,\R,\r)}:{-umax(\tdplotmaintheta,\R,\r)},smooth,samples=51]
 ({torusx(\x,vcrit2(\x,\tdplotmaintheta),\R,\r)},
 {torusy(\x,vcrit2(\x,\tdplotmaintheta),\R,\r)},
 {torusz(\x,vcrit2(\x,\tdplotmaintheta),\R,\r)})
 plot[variable=\x,
 domain={umax(\tdplotmaintheta,\R,\r)}:{180-umax(\tdplotmaintheta,\R,\r)},smooth,samples=51]
 ({torusx(\x,vcrit2(\x,\tdplotmaintheta),\R,\r)},
 {torusy(\x,vcrit2(\x,\tdplotmaintheta),\R,\r)},
 {torusz(\x,vcrit2(\x,\tdplotmaintheta),\R,\r)});
 \draw[thick] plot[variable=\x,
 domain={-180+umax(\tdplotmaintheta,\R,\r)/2}:{-umax(\tdplotmaintheta,\R,\r)/2},smooth,samples=51]
 ({torusx(\x,vcrit2(\x,\tdplotmaintheta),\R,\r)},
 {torusy(\x,vcrit2(\x,\tdplotmaintheta),\R,\r)},
 {torusz(\x,vcrit2(\x,\tdplotmaintheta),\R,\r)});
 \foreach \X  in {260}  
 {\draw[thick,dashed,color=DESYO] 
  plot[smooth,variable=\x,domain={360+vcrit1(\X,\tdplotmaintheta)}:{vcrit2(\X,\tdplotmaintheta)},samples=71]   
 ({torusx(\X,\x,\R,\r)},{torusy(\X,\x,\R,\r)},{torusz(\X,\x,\R,\r)});
 \draw[thick,dashed,color=DESYO] 
  plot[smooth,variable=\x,domain={vcrit2(\X,\tdplotmaintheta)}:{vcrit1(\X,\tdplotmaintheta)},samples=71]   
 ({torusx(\X,\x,\R,\r)},{torusy(\X,\x,\R,\r)},{torusz(\X,\x,\R,\r)});
 }
 \foreach \X  in {180}  
 {\draw[thick,dashed,color=DESYB] 
  plot[smooth,variable=\x,domain={0}:{360},samples=71]   
 ({torusx(\x,\X,\R+1,\r)},{torusy(\x,\X,\R+1,\r)},{torusz(\x,\X,\R+1,\r)});
}
\end{scope}

\begin{scope}[tdplot_main_coords,scale=0.5,xshift=-12cm,yshift=-3.8cm]
 \draw[thick,fill=gray,even odd rule,fill opacity=0.2] plot[variable=\x,domain=0:360,smooth,samples=71]
 ({torusx(\x,vcrit1(\x,\tdplotmaintheta),\R,\r)},
 {torusy(\x,vcrit1(\x,\tdplotmaintheta),\R,\r)},
 {torusz(\x,vcrit1(\x,\tdplotmaintheta),\R,\r)}) 
 plot[variable=\x,
 domain={-180+umax(\tdplotmaintheta,\R,\r)}:{-umax(\tdplotmaintheta,\R,\r)},smooth,samples=51]
 ({torusx(\x,vcrit2(\x,\tdplotmaintheta),\R,\r)},
 {torusy(\x,vcrit2(\x,\tdplotmaintheta),\R,\r)},
 {torusz(\x,vcrit2(\x,\tdplotmaintheta),\R,\r)})
 plot[variable=\x,
 domain={umax(\tdplotmaintheta,\R,\r)}:{180-umax(\tdplotmaintheta,\R,\r)},smooth,samples=51]
 ({torusx(\x,vcrit2(\x,\tdplotmaintheta),\R,\r)},
 {torusy(\x,vcrit2(\x,\tdplotmaintheta),\R,\r)},
 {torusz(\x,vcrit2(\x,\tdplotmaintheta),\R,\r)});
 \draw[thick] plot[variable=\x,
 domain={-180+umax(\tdplotmaintheta,\R,\r)/2}:{-umax(\tdplotmaintheta,\R,\r)/2},smooth,samples=51]
 ({torusx(\x,vcrit2(\x,\tdplotmaintheta),\R,\r)},
 {torusy(\x,vcrit2(\x,\tdplotmaintheta),\R,\r)},
 {torusz(\x,vcrit2(\x,\tdplotmaintheta),\R,\r)});
 \foreach \X  in {260}  
 {\draw[thick,dashed,color=DESYB] 
  plot[smooth,variable=\x,domain={360+vcrit1(\X,\tdplotmaintheta)}:{vcrit2(\X,\tdplotmaintheta)},samples=71]   
 ({torusx(\X,\x,\R,\r)},{torusy(\X,\x,\R,\r)},{torusz(\X,\x,\R,\r)});
 \draw[thick,dashed,color=DESYB] 
  plot[smooth,variable=\x,domain={vcrit2(\X,\tdplotmaintheta)}:{vcrit1(\X,\tdplotmaintheta)},samples=71]   
 ({torusx(\X,\x,\R,\r)},{torusy(\X,\x,\R,\r)},{torusz(\X,\x,\R,\r)});
 }
 \foreach \X  in {180}  
 {\draw[thick,dashed,color=DESYO] 
  plot[smooth,variable=\x,domain={0}:{360},samples=71]   
 ({torusx(\x,\X,\R+1,\r)},{torusy(\x,\X,\R+1,\r)},{torusz(\x,\X,\R+1,\r)});
}
\end{scope}

{
\begin{scope} [xshift=0.5cm, yshift=-2.6cm,scale=2,rotate around={58:(1.4,1.4)}]
\foreach \p in {(0.8,1.6)}
\fill[DESYB] \p circle(.1);

\draw[thick,fill=DESYB!30] (0.7,2.8) to [out=-60,in=-120]  (2.1,2.8)
 to [out=-90, in=180] (2.8,2.1) to [out=210, in=150] (2.8,0.7) to [out=180, in=90 ] (2.1,0) to [out=120, in=60]  (0.7,0) to [out=90, in=0] (0,0.7) to [out=30, in=-30] (0,2.1) to [out=0, in=-90] (0.7,2.8);
\end{scope}
}

\node at (-6,0) {\scalebox{3}{$\mathbf{+}$}};
\node at (-6,-3.8) {\scalebox{3}{$\mathbf{+}$}};
\node at (-6,-4.6) {\scalebox{3}{$\mathbf{\hdots}$}};
\node at (3.4,0.1) {\scalebox{3}{$\mathcal{M}_{Q,\alpha}$}};


\node[] (M3) at (-4,0) {}; 
\node[] (M4) at (-1,0) {}; 
\node[] (M1) at (-2,0) {}; 
\node[] (M2) at (0,0) {}; 

\draw [line width=1mm, -{Stealth[length=10mm, open]}] (M1) -- (M2);
\draw [line width=1mm, -{Stealth[length=10mm, open]}] (M4) -- (M3);
\foreach \p in {(-5,1.3),(-5,-2.5)}
\fill[black] \p circle(.1);
\node at (-4.8,1.1) {$\mathbf{\alpha}$};
\node at (-4.8,-2.7) {$\mathbf{\alpha}$};

\node[] (Y1) at (-6,4.3) {\scalebox{1.5}{\textbf{Sum Over}}};
\node[] (Y2) at (-6,3.6) {\scalebox{1.5}{\textbf{3D Geometries}}};

\node[] (P1) at (3.4,4.3) {\scalebox{1.5}{\textbf{Average Over}}};
\node[] (P2) at (3.4,3.6) {\scalebox{1.5}{\textbf{2D Moduli}}};

\end{tikzpicture}
}
\caption{Diagrammatic depiction of the holographic duality. On the bulk side, one performs a sum over geometries. Each geometry corresponds to a different choice of contractible cycle. Illustrated above are two examples of filling in the spatial (\textcolor{DESYO}{orange}) and temporal (\textcolor{DESYB}{blue}) cycles, which correspond to thermal AdS$_3$ and the BTZ black hole, respectively. Non-trivial Wilson lines wrap the non-contractible cycle and correspond to the worldlines of anyons labeled by $\alpha$. On the boundary side, one performs an average over (a cover of) the moduli space of the CFT.}
\label{fig:duality}
\end{figure}

\section{Emergence of Global Symmetries in the Boundary CFT}

We can now describe the emergence of a global symmetry after averaging generalized Narain CFTs. 
By definition, a zero-form symmetry $\sigma$ preserves the anyon data $(\mathscr{D}, \theta)$ up to a permutation of anyon labels. Since the anyon data determines the modular $S, T$-matrices \cref{eq:ST} and hence the ensemble-averaged partition function~\cref{eq:E_as_sum} due to the relation of $\gamma_\alpha^Q(c,d)$ with~\cref{lambda_def}, we find that Eisenstein series are simply permuted:
\begin{equation}
      E^{Q}_{\sigma\cdot \alpha}(\tau,\overline{\tau})=E^{Q}_{{\alpha}}(\tau,\overline{\tau}) 
      \fstop
      \label{eise}
\end{equation}
Of course, this is expected since the zero-form symmetries are also the global symmetries of the boundary theory.
The question we now address is to how these symmetries arise in the process of ensemble averaging. To do so, we examine the action of zero-form symmetries on the pre-averaged partition functions of~\cref{eq:lattheta} under the action $\alpha\rightarrow \sigma\cdot\alpha$.

When the zero-form symmetry is a quantum symmetry, the symmetry action changes the functional form of the theta function, and the symmetry is not simply a permutation. Quantum symmetries are then truly emergent and appear only after ensemble averaging as in~\cref{eise}.

By contrast, for a classical zero-form symmetry $\Sigma  \in \rmO_Q(p,q; \mathbb{Z})$, the symmetry preserves the CFT partition function up to an action on the moduli~\footnote{The action on the moduli $\Sigma\cdot m$ is defined as the replacement $H\rightarrow H^\prime := (\Sigma^{-1})^TH\Sigma^{-1}$ in~\cref{eq:lattheta}. Note that $H^\prime Q^{-1}H^\prime = Q$ for all $\Sigma\in \rmO_Q(p,q; \mathbb{Z})$.}:
\begin{equation}
  \vartheta^Q_{\Sigma\cdot \alpha}(\tau,\overline{\tau};\Sigma\cdot m) = \vartheta^Q_{\alpha}(\tau,\overline{\tau};m)\fstop
    \label{eq:thetaemsym}
\end{equation}
It is straightforward to check this statement explicitly.
More physically,
this essentially follows from the fact that the CFT moduli space~\cref{eq:M_Q} is already quotiented by the action of the T-duality group
$\rmO_Q(p,q; \mathbb{Z})$ \cref{eq:O_Q}. Note that before the ensemble average, the T-duality group acts both on the anyon label and the moduli,
while after the average it acts only on the anyon label. Thus the T-duality group of the original theory is ``folded'' into an emergent global symmetry via ensemble averaging---a process we call \textit{duality origami}.

While we discussed the case of generalized Narain theories, 
we can formulate this folding in general.
Let us consider the ensemble average of an observable $\mathcal{O}(m,x)$
of a theory $\mathcal{T}(m)$ (CFT or otherwise) over a moduli space $m\in \mathcal{M}$ \footnote{Note that the parameters $m, x$ play very different roles.
On the one hand, $m$ denotes the parameters for the specification of the 
theory inside the ensemble, and theories with different values of $m$ correspond to different theories.
On the other hand, the parameters $x$ will denote the parameters (such as coupling constants and mass parameters) of a single theory $\mathcal{T}(m)$ specified by $m$, and hence will not be averaged and will remain after the average.}.
The ensemble average of $\mathcal{O}(m,x)$ is defined as
\begin{align}\label{general_average}
\langle \mathcal{O}\rangle (x) : =\frac{1}{\textrm{Vol}(\mathcal{M}) } \int_{\mathcal{M}} [dm]\,  \mathcal{O}(m,x) \fcomma
\end{align}
where $[dm]$ is an appropriate measure of the moduli space.

If we assume that
(1) there exists a symmetry $G$ which preserves the moduli space $\mathcal{M}$
as well as its measure ($[d(g \cdot m) ] = [dm ]$) and that (2) $G$ acts covariantly on the observable $\mathcal{O}$ as
$\mathcal{O}(g\cdot m, g \cdot x) = \mathcal{O}( m,  x)$, 
then we call $G$ an {\it ensemble symmetry}.
Note that an ensemble symmetry is {\it not} a symmetry of a boundary theory
in the standard sense: an element $g\in G$ maps one theory (at a point $m$ in the moduli space) to another (at $g\cdot m$), 
and hence it relates two different theories inside the same ensemble.

Once we have an ensemble symmetry,  one can easily derive
\begin{align}
\langle \mathcal{O} \rangle (g\cdot x)= \langle \mathcal{O}\rangle (x)  \quad \textrm{ for} \quad g \in G
\end{align}
from the two conditions outlined above.
Since the dependence of the moduli $m$ is now removed,
we find a global symmetry acting on a 
single theory---this global symmetry is an emergent symmetry after the ensemble average.

We can formulate this as a general lesson in ensemble averages:
a symmetry connecting different theories in an ensemble
can be turned into an emergent global symmetry of a single theory after averaging.
 This is an interesting loophole to the holographic argument \cite{Harlow:2018jwu,Harlow:2018tng}
that there are no global symmetries in theories of quantum gravity. 

\section{Emergence of Global Symmetries in the Bulk}

Let us next discuss the emergence of global symmetries in the holographic bulk. As reviewed above, the bulk dual to the average of a Narain ensemble is an exotic theory of quantum gravity defined as a sum over geometries that each support an Abelian Chern-Simons theory. However, prior to the ensemble average, the bulk dual of a particular member of the ensemble is related to Maxwell-Chern-Simons theory \cite{Gukov:2004id,Ashwinkumar:2021kav}
\begin{equation}\label{eq:mym}
   S_{\rm MCS}= \frac{1}{16\pi^2}\sum_{i,j=1}^{p+q} \int_M \left(-\frac{1}{2e^2}\lambda_{ij}^{-1} d A^i\wedge * d A^j \right)
   + S_{\rm CS}
\fcomma
\end{equation}
where $e^2$ is the coupling that has dimensions of mass, and $\lambda^{-1}$ is a dimensionless, symmetric, positive definite matrix with a determinant one. Given that $e^2$ is dimensionful, the Maxwell term is 
irrelevant and therefore the Chern-Simons term is expected to dominate in the IR. This corresponds to the 
topological limit $e^2\rightarrow \infty$, which leaves only the Chern-Simons term.
The effect of the Maxwell term, however, still remains, since the quantization conditions for the gauge fields in the topological limit depend on the parameters $\lambda$, and hence on a point of the corresponding moduli space. Note that this duality involves a fixed bulk topology, and is an extension of the traditional $3D$ Chern-Simons/rational CFT correspondence~\cite{Witten:1988hf}, as discussed in~\cite{Gukov:2004id}.

As pointed out in \cite{Ashwinkumar:2021kav}, with standard boundary conditions imposed, the wavefunction of Maxwell-Chern-Simons theory matches the partition function of an irrational CFT with dependence on Narain moduli space (up to an overall constant) \cite{Gukov:2004id}, thereby realizing the holography duality prior to averaging. The same 
T-duality symmetries described earlier still exist for the Maxwell-Chern-Simons theory, where the mapping between different points in moduli space corresponds to a mapping between different points in theory space for the Maxwell-Chern-Simons theory. 

Maxwell-Chern-Simons theory of~\cref{eq:mym} is believed to describe the long-distance limit of   
 string theory on $AdS_3 \times K_7$, where $K_7$ is a compact 7-manifold~\cite{Gukov:2004id} \footnote{The perturbative partition function of tensionless string theory around $AdS_3\times S^3 \times T^4$ was studied in \cite{Eberhardt:2021jvj}, where it was shown that large stringy corrections admit an interpretation in terms of various semi-classical geometries. }. 
Although the gauge fields of Maxwell-Chern-Simons theory couple to other degrees of freedom in the corresponding low energy supergravity description, it was conjectured in \cite{Gukov:2004id}, based on the decoupling of topological modes at long distance, that the complete partition of string theory on $AdS_3 \times K_7$ is described by a linear combination of the 
partition functions $\vartheta_{\alpha}^Q$, which are now regarded as the wavefunction of Maxwell-Chern-Simons theory in the topological limit. Our discussion of holography is therefore relevant not only for quantum gravities in three spacetime dimensions, 
but also for the ten-dimensional string theory.

\section{Emergent Global Symmetries and Swampland Distance Conjecture}

Once we are in string theory, there is an alternative method to obtain global symmetries: going to the infinity of the moduli space.
Indeed, the Swampland distance conjecture \cite{Ooguri:2006in} states that there is a tower of states
in the infinite distance limit, where we often expect an emergence of a global symmetry \cite{Grimm:2018ohb,Heidenreich:2018kpg}.

In our situation, the natural limit for the modulus $\tau$
is to choose $\tau\to i\infty$, or its $SL(2, \mathbb{Z})$ images;
these are the cusps for the fundamental region of the torus (as in~\cref{eq:E_as_sum}), and 
are in the infinite distance limit of the moduli space (the parameter $\tau$ appears here as the complex structure parameter for the torus boundary of the $AdS_3$ subspace of the $AdS_3 \times K_7$ geometry in the bulk). Each cusp corresponds to a geometry \cite{Ashwinkumar:2021kav},
and in the infinite distance limit to the cusp, one of the cycles of the associated geometries shrinks to zero size (so that three-dimensional gravity reduces to two-dimensional gravity).
As discussed above, the leading divergence at the cusp is given by the lens space partition function, which is the quantity associated with the 
post-averaged bulk theory and encapsulates the global symmetry. Note that this infinite distance limit is distinct from the usual Swampland distance conjecture, which deals with displacements of compactification moduli~\cite{Ooguri:2006in}. In contrast, our limit is one on the geometry of the AdS$_3$ spacetime, and is more akin to the Generalized Distance Conjecture of~\cite{Lust:2019zwm}.

It is interesting to compare the two different types of emergence of global symmetries discussed above.
One is obtained by taking an infinite-distance limit of a modulus and is associated with the Swampland distance conjecture.
Another is obtained by ensemble averaging of the CFT moduli space.
In both cases, we are led to exactly the same expression, namely the lens space partition function. This discussion is reminiscent of the results of~\cite{Collier:2022emf}, where the ensemble average maps one to the large $N$, large 't Hooft coupling limit of the $\mathcal{N}=4$ Super Yang-Mills boundary CFT. These results hint towards deeper connections between ensemble averages in holography, the Swampland program, and string theory.

\section{Fluctuations around Ensemble Averages and Breaking Emergent Symmetries}

When we take seriously the correspondence between ensemble averaging and distance conjecture, a natural question is how the global symmetries are broken in honest theories of quantum gravity.
In the case of the distance conjecture, this is achieved by staying at a finite distance in the moduli space.

In analogy with this, one could re-introduce moduli dependence in the ensemble average picture as well. To make sense of this, one should take the viewpoint that ensemble-averaged based holographic correspondences are ``coarse-grained" descriptions of microscopic dualities. By re-introducing moduli dependence, one is going from a coarse-grained to a more fine-grained picture. The moduli dependence can then be interpreted as deviations or fluctuations from the ensemble-average, and should be responsible for removing any emergent global symmetries. One can see that this is so in the discussions above - the theta functions, which depend on moduli, either do not respect the emergent symmetries or transmute them into T-duality transformations. However, in the expressions above, it is not clear how to interpret the theta functions as corresponding to an ensemble average with corrections. 
The question is then 
how to formulate these ideas in a precise mathematical formalism.

Remarkably, the relevant mathematics is already known in the literature
as the Roelcke-Selberg spectral decomposition. This is a decomposition of a square-integrable modular form as in~\cite{Terras,Zagier:1981}. This technology was applied to the partition functions of Narain CFTs defined via even, self-dual lattices in~\cite{Benjamin:2021ygh}, where it was shown that the ensemble average arises as the moduli-independent piece of the spectral decomposition. For our generalized CFTs, we require a spectral decomposition for non-holomorphic modular forms of congruence subgroups with non-zero weight. For square-integrable forms $f(\tau,\taub)$, the decomposition is~\cite{roelcke1966eigenwertproblem,roelcke1967eigenwertproblem,mono2019spectral} 
\begin{align}\label{eq:f_decomposition}
&f(\tau,\taub)=  \sum_{i=1}^{\infty}\left\langle f, u_i\right\rangle u_i(\tau)+\sum_{j=1}^N\left\langle f, v_j\right\rangle v_j(\tau)\nonumber \\
& +\sum_{k=1}^n \frac{1}{4 \pi} \int_{-\infty}^{\infty}\left\langle f, E_{\mathfrak{a}_k}\!\left(\tau, \frac{1}{2}+i t\right)\right\rangle E_{\mathfrak{a}_k}\!\left(\tau, \frac{1}{2}+i t\right) d t\fstop
\end{align}
Here $\{u_i\}$ is the orthonormal basis of cusp forms, $\{v_j\}$ the residues of the Eisenstein series,  $E_{\mathfrak{a}_k}$ are Eisenstein series labeled by the cusps $\mathfrak{a}_k$ of the relevant congruence subgroup, and $\langle -, -\rangle$ is the Petersson inner product; see \cite[Theorem 6.7.1]{mono2019spectral} for more information.
When we apply this decomposition to the combination
measuring the deviation from the ensemble average
\ie \label{eq:f_choice}
f(\tau,\bar{\tau})=\tau_2^{(p+q)/4}(\vartheta_{Q,h}(\tau, \bar{\tau} ; m)-E_{Q, h}(\tau,\bar{\tau}))\fcomma
\fe 
the terms in the decomposition are all moduli-dependent and represent deviations from the ensemble average, hence triggering the breaking of the 
the emergence of global symmetries. 
This suggests that the counterparts of the ``tower of states'' in the Swampland distance conjecture
should be included inside the moduli-dependent terms of the spectral decomposition. 
We also expect that such an analysis will be related to the discussion of wormholes before averaging in \cite{Saad:2021rcu}.
It is an interesting question to study this point further.

One of the surprises of quantum gravity and string theory is that they have successfully incorporated
a wide variety of ideas. It is therefore natural to expect that
Swampland and ensembles, which at first seem to be in tension with each other, are combined nicely in string theory,
and that such a combination will lead us to deeper insights into the mysteries of quantum gravity~\footnote{See~\cite{Heckman:2021vzx} for an alternative discussion on embedding ensemble averaging in string theory, as well as its relation to the Hilbert space of baby universes in~\cite{McNamara:2020uza, Marolf:2020xie}}.\\

After this work was completed, we learned of~\cite{Antinucci:2023uzq}, which has related discussions on the topic of global symmetries and ensemble averaging.

\section{Acknowledgments}
We thank A.\ Kidambi, C.\ Nazaroglu, A.\ Westphal, and Y.\ Zheng for useful discussions. 
M.Y.\ is grateful to the KITP, Santa Barbara for hospitality during the Integrable22 workshop.
M.A.\ and M.Y.\ are supported in part by the JSPS Grant-in-Aid for Scientific Research (19H00689, 20H05860).
J.M.L. is supported by the Deutsche Forschungsgemeinschaft under Germany's Excellence Strategy - EXC 2121 ``Quantum Universe'' and Deutsche Forschungsgemeinschaft through a German-Israeli Project Cooperation (DIP) grant “Holography and the Swampland”. J.M.L. thanks Kavli Institute for the Physics and Mathematics of the Universe for its hospitality during the completion of a portion of this work. M.Y.\ is also supported in part by the JSPS Grant-in-Aid for Scientific Research (19K03820, 23H01168), and by JST, Japan (PRESTO Grant No.\ JPMJPR225A, Moonshot R\&D Grant No.\ JPMJMS2061).

\bibliography{EmergentSym}

\end{document}